\documentclass[graybox, envcountchap]{svmult}

\usepackage{mathptmx}        
\usepackage{amsmath}
\usepackage{amssymb}
\usepackage{color}
\usepackage{helvet}          
\usepackage{courier}         
\usepackage{dirtree}

\usepackage{makeidx}        
\usepackage{graphicx}        
\usepackage{subfig}

\usepackage{multicol}        
\usepackage[bottom]{footmisc}

\usepackage{hyperref}        
\hypersetup{colorlinks=true,urlcolor=blue,citecolor=blue}

\usepackage[misc]{ifsym}

\usepackage{epstopdf}
\usepackage{bm}
\usepackage{upgreek}
\usepackage[
    backend=biber,
    sorting=nty,
    maxnames=1, minnames=1, firstinits=true, terseinits=true,
    citestyle=numeric,
    bibstyle=authoryear,
]{biblatex}

\makeatletter
\input{numeric.bbx}
\makeatother

\DeclareNameAlias{author}{first-last}
\DeclareFieldFormat*{title}{}
\renewbibmacro{in:}{}
\DeclareFieldFormat{pages}{#1}  

\include{journals}

\makeindex             

\begin{document}

\newcommand{\ho}{$H_0$}
\newcommand{\Ho}{$H_0$}
\newcommand{\hunit}{km s$^{-1}$ Mpc$^{-1}$}
\newcommand{\examplelens}{RXJ1131$-$1231}
\newcommand{\macs}{MACS1149.5$+$2223}


\title{Strong Lensing and \Ho}
\author{Tommaso Treu \& Anowar J.~Shajib}
\institute{Tommaso Treu (\Letter) \at Physics \& Astronomy Department, University of California, Los Angeles, CA, 90095, USA, \email{tt@astro.ucla.edu}
\and Anowar J.~Shajib \at NHFP Einstein Fellow, Department  of  Astronomy  \&  Astrophysics,  University  of Chicago, Chicago, IL 60637, USA, \email{ajshajib@uchicago.edu} \\
Kavli Institute for Cosmological Physics, University of Chicago, Chicago, IL 60637, USA
}
%
%
\maketitle

\abstract{Time delays from strong gravitational lensing provide a one-step absolute distance measurement. Thus, they measure \Ho\ independently of all other probes. We first review the foundations and history of time-delay cosmography. Then, we illustrate the current state of the art by means of two recent case studies that have been real breakthroughs: i) the quadruply imaged quasar lensed by a galaxy-scale deflector \examplelens, for which spatially resolved stellar kinematics is available; ii) the multiply imaged supernova ``Refsdal'', the first with measured time delays, lensed by cluster \macs. We conclude by discussing the exciting future prospects of time-delay cosmography in the coming decade.}


\section{Introduction}
\label{sec:intro}

Following Fermat's Principle, multiple images of the lensed source form at stationary points of the arrival time surface. The difference in arrival time between multiple images arises from the combination of the geometric delay, which makes some light paths longer than others, and the Shapiro delay \cite{Shapiro1964}, arising from the difference in gravitational potential along different paths.

Thus, if one can measure the difference in arrival time and reconstruct the gravitational potential, one obtains an absolute measurement of the difference in the length of the light paths. It is then a simple matter to convert this measurement into a combination of angular diameter distances, known as the ``time-delay distance'' \cite{Suyu10}, and thus obtain a direct measurement of the Hubble constant, \ho\ \cite{Refsdal64}.

The principle is elegant and straightforward, and Refsdal \cite{Refsdal64} recognized its potential well before strong gravitational lenses were discovered in the late 70s \cite{Walsh79}. Since then, breakthroughs in observations, methodology, and theory have finally made Resfdal's dream a reality. Time-delay cosmography, as it has come to be called, has been demonstrated to yield measurements of \ho\ at the level of precision and accuracy of a few percent for a single strong lens system \cite{Suyu10, Shajib19}. As strong lens systems are being discovered and followed up at an increasing pace, the overall precision achievable by combining multiple systems is improving rapidly. Progress has also been achieved in understanding systematic errors \cite{Millon20, Gilman20, Birrer20, VandeVyvere22, Gomer22, Ding21b} and how to control them so as to achieve unbiased sample averages at the level of 1-2\% in accuracy and precision required to help settle the Hubble tension \cite{Birrer21}.

Time-delay cosmography has several advantages as a probe of \Ho. This method provides a one-step measurement of \Ho\ that is completely independent of other \Ho\ probes. The measured cosmological distances are angular diameter distances. Thus, the method is not susceptible to uncertain dust extinction laws that luminosity distance indicators could be susceptible to \cite{Efstathiou2021}. 

In this Chapter, we first provide some background in the theory and history of the method in Section~\ref{sec:background}. We then review the current state of the art in Section~\ref{sec:art}, by covering two recent case studies in some detail. First, in ~\ref{sec:time_delay_measurement} to Section~\ref{sec:stellar_kinematics}, we describe the case of \examplelens, a galaxy deflector lensing a quasar, that has been modeled based on excellent data, including ground-based spatially resolved kinematics. Second, in Section~\ref{sec:sn} we described the case of supernova Refsdal, multiply imaged by a foreground cluster, the first lensed supernova that has been used to measure H$_0$. We conclude with our future outlook in Section~\ref{sec:outlook}.
Due to space limitations, we only focus on the main points currently relevant to the measurement of \ho\ and refer to previous reviews for more extensive treatments of the theory and history of the method \cite{Treu16,Suyu2018,Treu22,Birrer23}.

\section{Background}
\label{sec:background}

\subsection{Theory}

This section briefly reviews the theory of time-delay cosmography. See \cite{Schneider92} for a detailed treatment of the strong lensing formalism.

\subsubsection{Strong lensing formalism}


The lensing phenomenon is described by the lens equation
\begin{eqnarray}
\bm{\beta} = \bm{\theta} - \bm{\alpha}(\bm{\theta})
\end{eqnarray}
 that maps a source plane coordinate vector $\bm{\beta}$ to an image plane coordinate vector $\bm{\theta}$, where $\bm{\alpha}(\bm{\theta})$ is the deflection angle vector. The lensing deflection is produced by the mass distribution between the source and the observer. In the thin lens approximation, the lensing mass distribution is described by the surface mass density $\Sigma(\vec{\theta})$ projected on the lens plane. The  dimensionless lensing convergence is defined as $\kappa \equiv \Sigma / \Sigma_{\rm cr}$, where the critical density $\Sigma_{\rm cr}$ is given by
\begin{eqnarray}
    \Sigma_{\rm cr} \equiv \frac{c^2}{4 \uppi G} \frac{D_{\rm s}}{D_{\rm d} D_{\rm ds}}.
\end{eqnarray}
Here, $D_{\rm d}$ is the angular diameter distance between the observer and the deflector, $D_{\rm s}$ is the angular diameter distance between the observer and the source, and $D_{\rm ds}$ is the angular diameter distance between the deflector and the source. The angular diameter distance between two redshifts $z_1$ and $z_2$ is given by
\begin{eqnarray}
    D_{\rm A} (z_1, z_2) = \frac{c}{H_0(1 + z_2)} f_k (z_1, z_2, \Theta),
\end{eqnarray}
where $\Theta$ is the set of cosmological parameters excluding \Ho\ in a given cosmology and $f_k(z_1, z_2, \Theta)$ is a function whose form depends on the sign of the curvature density $\Omega_k$ \cite{Weinberg72, Peebles93}.

The deflection angle $\bm{\alpha}$ is related to the convergence $\kappa$ as
\begin{eqnarray}
    \kappa = \frac{1}{2} \nabla \cdot \bm{\alpha},
\end{eqnarray}
and to the lensing potential $\psi$ as
\begin{eqnarray}
    \bm{\alpha} (\bm{\theta}) = \nabla \psi (\bm{\theta}).
\end{eqnarray}

Thus, the lensing potential is connected to the surface mass distribution by the two-dimensional Poisson equation:

\begin{eqnarray}
    \nabla^2 \psi (\bm{\theta}) =  2 \kappa (\bm{\theta}).
\end{eqnarray}

We can define the Fermat potential \cite{Schneider85, Blandford86} as
\begin{eqnarray}
    \tau (\bm{\theta}, \bm{\beta}) = \frac{(\bm{\theta} - \bm{\beta})^2}{2} + \psi (\bm{\theta}).
\end{eqnarray}
The time delay between photon arrival times between images A and B is given by
\begin{eqnarray} \label{eq:time_delay}
    \Delta t = \frac{D_{\Delta t}}{c} \left[ \tau (\bm{\theta}_{\rm A}) - \tau (\bm{\theta}_{\rm B}) \right],
\end{eqnarray}
where $D_{\Delta t}$ is the ``time-delay distance'' \cite{Refsdal64, Suyu10} defined as
\begin{eqnarray} \label{eq:time_delay_distance}
    D_{\Delta t} \equiv (1 + z_{\rm d}) \frac{D_{\rm d} D_{\rm s}}{D_{\rm ds}}.
\end{eqnarray}
The time delay $\Delta t$ can be measured for transient sources such as supernovae (SNe) or quasars. The time-delay distance is inversely proportional to the Hubble constant $H_0$. Therefore, if we can measure the time delay $\Delta t$ and the lensing potential $\psi$, we can measure $H_0$. The two unknowns $\Delta \psi$ and $\bm{\beta}$ in equation \ref{eq:time_delay} are obtained by modeling the data. The practical aspects of lens modeling with imaging data are described in Section \ref{sec:lens_modeling}. 
The next section discusses the well-known ``mass-sheet degeneracy'' \cite[MSD][]{Falco85, Schneider92}. Limiting the MSD is crucial to achieving precise and accurate H$_0$ measurements.

\subsubsection{Mass-sheet degeneracy}

The mass-sheet transformation (MST) of the convergence $\kappa$ given by
\begin{eqnarray}
    \kappa \to \kappa' = \lambda \kappa + (1 - \lambda)
\end{eqnarray}
leaves all the imaging observables invariant with a simultaneous rescaling of the unknown source position 
\begin{eqnarray}
	\bm{\beta} \to \bm{\beta}^{\prime} = \lambda \bm{\beta}, 
\end{eqnarray}
where $\lambda$ is a constant. This degeneracy in the convergence $\kappa$ and thus the potential $\psi$  is called the mass-sheet degeneracy (MSD).

The time delay $\Delta t$ transforms under the mass-sheet transformation as
\begin{eqnarray}
    \Delta t \to \Delta t' = \lambda \Delta t.
\end{eqnarray}
Thus, the time-delay distance $D_{\Delta t}$ and  Hubble constant $H_0$ inferred from the observed time-delay $\Delta t$ will change as
\begin{eqnarray}
    D_{\Delta t}^{\prime} = D_{\Delta t} / \lambda, \\
	H_0^{\prime} = \lambda H_0. 
\end{eqnarray}

To gain an understanding of the MST, it is useful to describe the ``true'' physical convergence $\kappa_{\rm true}$ in terms of two components as
\begin{eqnarray}
    \kappa_{\rm true}(\bm{\theta}) = \kappa_{\rm int}(\bm{\theta}) + \kappa_{\rm ext},
\end{eqnarray}
where $\kappa_{\rm int}(\bm{\theta})$ is the convergence produced by the mass distribution of the galaxy or group or cluster acting as the main deflector, while $\kappa_{\rm ext}$ is the convergence produced by the mass distribution not physically associated with the main deflector, e.g., along the line of sight (LOS). Since for a physical deflector $\lim_{\bm{\theta} \to \infty} \kappa_{\rm int}(\bm{\theta}) = 0$, we can see that $\lim_{\bm{\theta} \to \infty} \kappa_{\rm true}(\bm{\theta}) = \kappa_{\rm ext}$. Due to the mass sheet degeneracy, this external convergence term cannot be constrained from the imaging of the lensing system. However, it can be estimated by comparing the statistics of LOS mass distribution between cosmological simulations and that observed using photometric and spectroscopic surveys (Section \ref{sec:los}). If the external convergence $\kappa_{\rm ext}$ is ignored during lens modeling, then the modeled convergence $\kappa_{\rm model}^{\prime}$ will be an MST of the true convergence $\kappa_{\rm true}$ with $\lambda = 1/(1 - \kappa_{\rm ext})$ as
\begin{eqnarray}
    \kappa_{\rm model}^{\prime} (\bm{\theta}) = \frac{\kappa_{\rm true}(\bm{\theta})}{1 - \kappa_{\rm ext}} + 1 - \frac{1}{1 - \kappa_{\rm ext}} = \frac{\kappa_{\rm int}(\bm{\theta})}{1 - \kappa_{\rm ext}}.
\end{eqnarray}
Here, the condition $\lim_{\bm{\theta} \to \infty} \kappa_{\rm model}^{\prime} = 0$ is satisfied, since $\kappa_{\rm model}^{\prime}$ is only attributed to the central galaxy's or galaxies' mass distribution that ought to vanish at infinity. 

In practice, lens models are often described with simply parameterized profiles, which implicitly break the MSD. Therefore, the best fit simply-parametrized model $\kappa_{\rm model}$ could be thought to be an approximate MST of $\kappa_{\rm model}^{\prime}$ as
\begin{eqnarray}
    \kappa_{\rm model}^\prime (\bm{\theta}) = \lambda_{\rm int}(\bm{\theta}) \kappa_{\rm model} (\bm{\theta}) + 1 - \lambda_{\rm int} (\bm{\theta}).
\end{eqnarray}
Here, the internal MST parameter $\lambda_{\rm int}(\theta)$ cannot be a constant mass-sheet to satisfy $\lim_{\bm{\theta} \to \infty} \kappa_{\rm model}^{\prime} = 0$. A $\lambda_{\rm int}(\bm{\theta)}$ can be designed, which acts as a constant sheet of mass near the central region of the lensing system ($\theta \lesssim \mathcal{O}(10\theta_{\rm E})$), but vanishes at $\theta \gg \theta_{\rm E}$ \cite{Blum20, Birrer20}, where $\theta_{\rm E}$ is the Einstein radius. As a result, the true internal mass distribution can be expressed as
\begin{eqnarray}
    \kappa_{\rm int} (\bm{\theta}) = (1 - \kappa_{\rm ext})[ \lambda_{\rm int}(\bm{\theta}) \kappa_{\rm model}(\bm{\theta}) + 1 - \lambda_{\rm int}(\bm{\theta})]
\end{eqnarray}
Thus, the true Fermat potential difference relates to the modeled Fermat potential difference as
\begin{eqnarray} \label{eq:fermat_potential_mst}
    \Delta {\tau}_{\rm true} =  \Delta {\tau}_{\rm model} \lambda_{\rm int} (1 - \kappa_{\rm ext}).
\end{eqnarray}
Here, we have expressed $\lambda_{\rm int}$ as a constant since it is approximately a constant in the region where lensed images appear.

\begin{svgraybox}
Combining equations \ref{eq:time_delay}, \ref{eq:time_delay_distance}, and \ref{eq:fermat_potential_mst}, the measured Hubble constant from time-delay cosmography can be expressed as
\begin{eqnarray} \label{eq:h0_master_equation}
    H_0 = \frac{ \Delta \tau_{\rm model} (1 - \kappa_{\rm ext}) \lambda_{\rm int}} {\Delta t} \frac{f_k ({0, z_{\rm d}, \Theta}) f_k(0, z_{\rm s}, \Theta)}{f_k (z_{\rm d}, z_{\rm s}, \Theta)}.
\end{eqnarray}
\end{svgraybox}
In this formulation, the time delay $\Delta t$ is directly measured from light curves of the images (Section \ref{sec:time_delay_measurement}), $\Delta \tau_{\rm model}$ is obtained from lens modeling of high-resolution imaging data of the lens system (Section \ref{sec:lens_modeling}), $\kappa_{\rm ext}$ is estimated from photometric and spectroscopic surveys of the lens environment (Section \ref{sec:los}), and $\lambda_{\rm int}$ is constrained from the stellar kinematics of the lens galaxy (Section \ref{sec:stellar_kinematics}). 

This formulation also illustrates that $H_0$ measured by time-delay cosmography weakly depends on the expansion history of the Universe from redshift $z_{\rm s}$, through redshift $z_{\rm d}$, up to redshift $z = 0$. Usually, a cosmological model is assumed to compute the $f_k$ functions, and $H_0$ can be slightly degenerate with other cosmological parameters, such as the dark energy's equation of state parameter $w$ \cite{Bonvin18, Wong20}. However, it is also possible to infer $H_0$ by constraining the Universe's expansion history empirically, using relative distances of the type Ia supernova. This inverse distance ladder method \cite{Taubenberger19} allows one to obtain \Ho\ without assuming specific values for the other cosmological parameters.

\subsection{A brief history}

After Refsdal's \cite{Refsdal64} suggestion, time-delay cosmography stayed dormant until the first lensed quasars were discovered some 15 years later \cite{Walsh79}. A period of excitement followed in the 1980s when astronomers tried to measure time delays for the first time \cite{Falco85}. However, determining time delays proved to be a significant challenge. First, the stochastic nature of quasar light curves makes it more difficult than using the well-behaved supernovae light curve, as originally suggested by Refsdal. Second, light curves at optical wavelengths (corresponding usually to rest frame UV or blue) are severely affected by microlensing, which further confounds the signal. Third, the typical image separation of galaxy-scale lenses is similar to the image quality of seeing-limited ground-based optical telescopes. 

Overcoming these challenges required monitoring campaigns with significantly higher precision per epoch, higher cadence, and longer duration than initially thought. By the end of the 1990s and early 2000s, such campaigns started to produce reliable time delays with few percent precision, both in the optical \cite{Kun++97, Sch++97,Oscoz1997} and in the radio \cite{Fassnacht02}. 

Once the ability to obtain time delays was demonstrated, attention turned to constraining the lensing potential. A two-pronged approach proved successful. On the one hand, improvements in data quality and lens modeling techniques allowed astronomers to capture the information content of multiply imaged extended sources, e.g., quasar host galaxies and radio jets and emissions. This step increases by several orders of magnitude the constraints on the mass distribution of the deflector with respect to the positions of the quasars in the image plane used in previous work. 

On the other hand, it became possible to measure the stellar kinematics of the deflector, a dynamical mass tracer \cite{T+K02b} that is crucially insensitive to the limiting factors of lensing, chiefly the MSD. Stellar kinematics also provides an independent handle on the angular diameter distance to the deflector, further enhancing the cosmological constraints \cite{Jee15, Jee16}. Conversely, lensing is insensitive to the mass-anisotropy degeneracy affecting dynamical estimates of mass. As shown in Section~\ref{sec:stellar_kinematics}, the combination of lensing and stellar dynamics is crucial to the success of this methodology~\cite{Shajib18}.

The final piece of the puzzle for precision time-delay cosmography at the galaxy scale was accounting for the effects of the line of sight and the local environment \cite{Suyu10}. Under or overdense lines of sight, nearby perturbers, multiple-plane lensing, and group/cluster scale halos can affect the inferred distances for galaxy-scale lenses as much as 5--10\% if not properly accounted for. 

Fittingly, 50 years after Refsdal's paper, the first multiply imaged supernova was discovered in 2014 \cite{Kelly15}. The supernova, properly named ``Refsdal'', was multiply imaged by a cluster of galaxies, making the geometry and arrival time sequence significantly more complex than those observed for quasars lensed by galaxy-scale potentials. The discovered ``Einstein-cross'' configuration is primarily due to a cluster galaxy and has little value for cosmography since the time delays are short \cite{Rodney2016}. Additional images, however, were predicted upon discovery based on the cluster mass model \cite{Kelly15,Oguri15,Sharon15}. One was in the past and thus sadly lost, but the other was predicted to re-appear approximately a year later \cite{Treu16b}, and it indeed appeared when and where the model predicted \cite{Kelly16}. Measurements of \Ho\ based on SN Refsdal have been reported \cite{Kelly2023a, Kelly2023b}. 

Additional multiply imaged supernovae have been discovered since SN Refsdal \cite{Frye2023}, and this is an area where much growth is expected in the coming decade.

\section{Current State of the Art}
\label{sec:art}

We now review the current state of the art by means of two case studies. First, in Sections~\ref{sec:time_delay_measurement} to Section~\ref{sec:stellar_kinematics}, we describe the galaxy-scale lens \examplelens. In this system, the variable source is a quasar (Fig.~\ref{fig:rxj1131_lens_model}), and it is arguably the galaxy-scale system with the highest quality dataset to date. Then, in Section~\ref{sec:sn}, we examine the case of the SN Refsdal, lensed by cluster \macs. This is the first example of \Ho\ measured via a multiply imaged supernova and presents a number of interesting differences with respect to the more established method of using quasars lensed by galaxies.


\subsection{Time delay measurements} \label{sec:time_delay_measurement}

\begin{figure*}
	\includegraphics[width=0.8\textwidth]{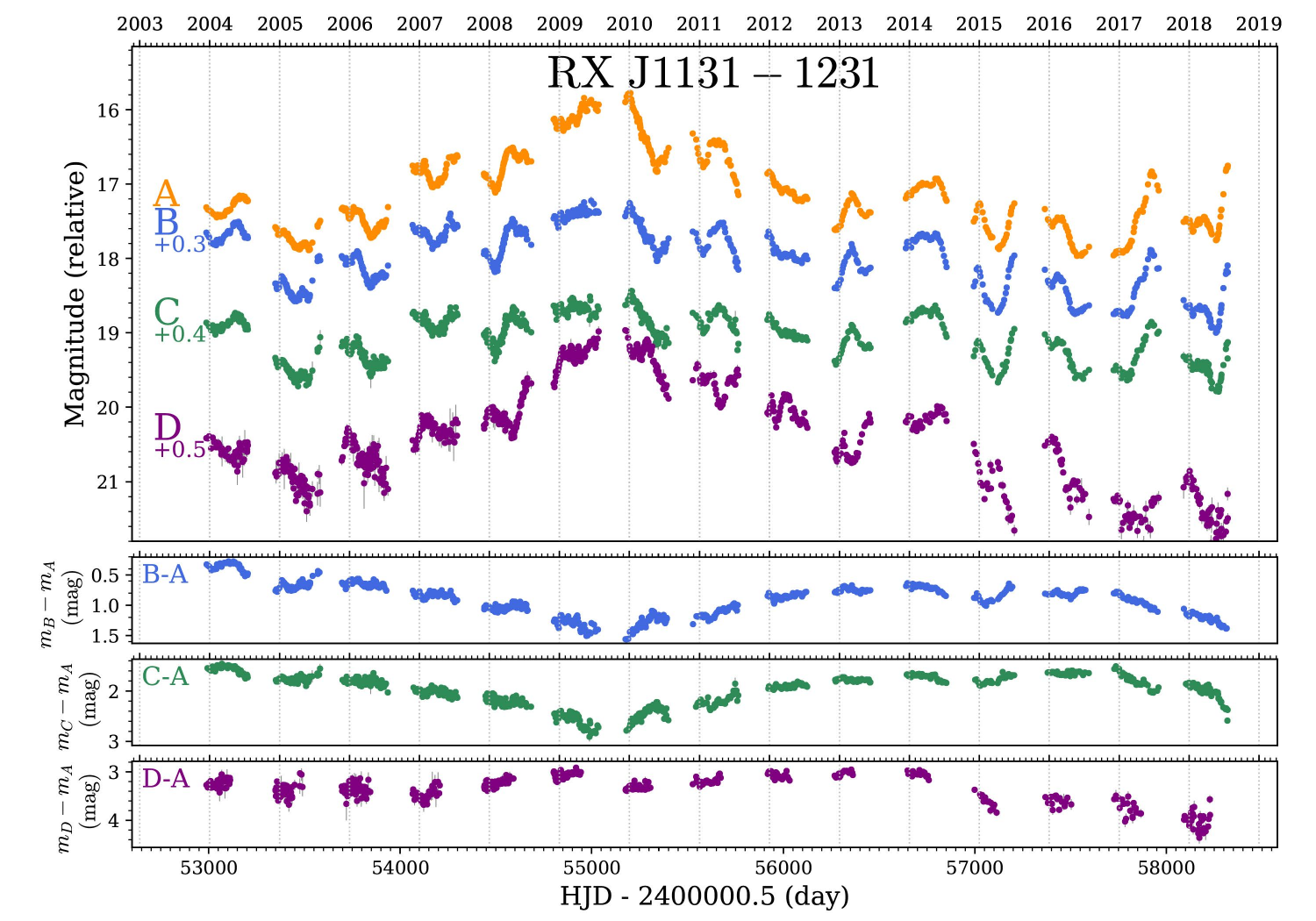}
	\caption{\label{fig:rxj1131_light_curve}
	Light curve of \examplelens\ in the $R$-band from COSMOGRAIL. The bottom panels show the difference curves between pairs of images after shifting the curves corresponding to the measured delays. These difference curves illustrate the long-term extrinsic variability due to microlensing. Figure from \cite{Millon20a}.
	}
\end{figure*}

Time delays (i.e., $\Delta t$ in Eq.~\ref{eq:h0_master_equation}) between lensed quasar images are measured using the quasar's intrinsic variability. The light curves of the individual images are measured with 1-m or 2-m class telescopes. The COSMOGRAIL collaboration \cite{Courbin05} has monitored several lensed quasar systems, spanning up to more than a decade, providing precise measurements to a few percent \cite{Courbin11, Tewes13, Millon20a}. Fig.~\ref{fig:rxj1131_light_curve} shows light curves from COSMOGRAIL for \examplelens\ based on 16 years (2003--2019) of monitoring. \cite{Courbin18} have demonstrated that robust time delays can be measured within a single season with almost a daily cadence and millimag photometric precision to detect small amplitude variations of the order of 10--50 millimag. Given the numerous discoveries of new lensed quasar systems from recent, ongoing, and future surveys, such fast turnaround for time delay measurements will be essential to rapidly produce cosmological constraints.

The main challenge in measuring the time delay is the extrinsic variability caused by the microlensing of foreground stars. This extrinsic variability pattern is unique to each lensed image. The microlensing variability can be of two types. The first one is a fast rise and fall, giving a sharp peak in the light curve. This fast variability happens when the lines of formally infinite magnification \cite[caustics][]{Schneider92} from a single foreground star cross over the quasar accretion disk. The second one is a long-term variability owing to overlapping caustics resulting from crowded stars creating a smooth change in the microlensing magnification as they move in front of the background quasar, due to internal and peculiar transverse motions.

Techniques developed to extract the time delays from the light curves while accounting for the microlensing variability include: cross-correlation that does not require explicit modeling of the microlensing variability \cite{Pelt96}; explicit modeling of the intrinsic and microlensing variabilities with spline fitting or Gaussian processes \cite{Tewes13, Hojjati13}. The ``Time Delay Challenge'' (TDC) \cite{Dobler15,Liao15} validated these techniques with simulated data. 
The \textsc{PyCS} software used for the COSMOGRAIL data analysis was among the techniques achieving the target precision and accuracy in the TDC. 

\subsection{Lens modeling} \label{sec:lens_modeling}

\begin{figure*}[!t]
	\includegraphics[width=\textwidth]{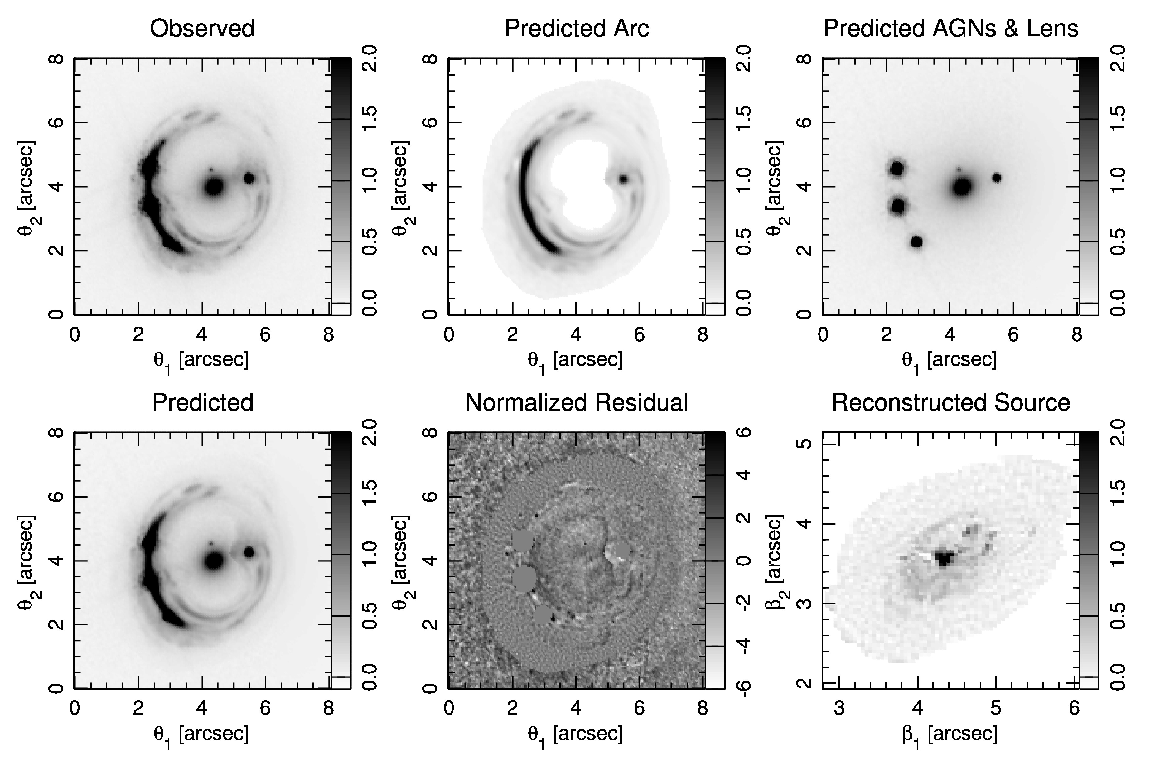}
	\caption{\label{fig:rxj1131_lens_model}
	Illustration of lens model of \examplelens\ from \cite{Suyu10}. \textbf{Top left:} Observed \textit{HST} ACS/F814W image. \textbf{Top middle:} Reconstructed lensed arcs of the quasar host galaxy using the best-fit lens model. \textbf{Top right:} Reconstructed quasar images and the lens galaxy light. \textbf{Bottom right:} Total reconstructed image obtained by summing the top-middle and top-right panels. \textbf{Bottom middle:} Normalized residuals for the reconstructed image. \textbf{Bottom right:} The reconstructed quasar host galaxy morphology on the source plane.
	}
\end{figure*}

The main objective in lens modeling is to constrain the lensing potential that gives rise to the observed lensed images and distorted arcs from the background quasar and its host. This lensing potential provides $\Delta \tau_{\rm model}$ in Eq.~\ref{eq:h0_master_equation}. Usually, high-resolution imaging from the \textit{HST} is used in lens modeling due to the superb stability in the point spread function (PSF) and diffraction-limited nature of the PSF to resolve lensed quasars that are separated by $\sim 1^{\prime\prime}$ (e.g., Figure \ref{fig:rxj1131_lens_model}). However, adaptive-optics-assisted imaging from large ground-based telescopes, such as the NIRC2 imager at the Keck Observatory, has also been successfully modeled for cosmographic analysis after meticulous reconstruction of the PSF \cite{Chen19}.

The lens model has four main components: (i) the lensing potential or the deflector mass distribution, (ii)  the flux distribution in the deflector galaxy, (ii) the flux distribution in the quasar host galaxy, and (iv) the point spread function (PSF). Thus, the model for the imaging data can be reconstructed from which the likelihood function $p(\bm{d}_{\rm imaging} \mid \xi_{\rm mass}, \xi_{\rm light}, \xi_{\rm source}, \mathcal{P})$ can be computed, where $\bm{d}_{\rm imaging}$ is the imaging data, $\xi_{\rm mass}$ is the mass model parameters, $\xi_{\rm light}$ is the deflector galaxy's light model parameter, $\xi_{\rm source}$ is the quasar host galaxy's light model parameter, and $\mathcal{P}$ is the PSF model. The PSF $\mathcal{P}$ can be initially estimated from a few stars within the imaging data. However, due to color mismatch between the stellar type and the quasar, spatial variations, and undersampling, this initial PSF model needs to be improved during the lens modeling to fit the pixels around the lensed quasar images to the noise level \cite{Chen16, Birrer19}. The PSF also usually needs to be reconstructed at a higher pixel resolution than the original image \cite{Shajib22, Shajib22b}. The lens model parameters $\xi_{\rm mass}$, $\xi_{\rm light}$, and $\xi_{\rm source}$ are then constrained by sampling the likelihood function.

The most common choices for the deflector galaxy's mass or potential model are ``simply parametrized'' models, where the mass distribution of the main deflector galaxy and other nearby galaxies are described with functions depending on a few free parameters. The nearby galaxies along the LOS are added to the parametric lens model when their higher order lensing effect (i.e., flexion) is non-negligible \cite{McCully17, Wong17, Birrer19b, Shajib20}. The lensing contribution from all the other LOS structures can be accounted for by the independently estimated external convergence term (Sec.~\ref{sec:los}) and an external shear component added to the lens model's parametric description. The simplest choice that yields good residuals for the simply parametrized model of galaxy-scale lenses is the elliptical power-law mass distribution with the 3D radial density profile $\rho(r) \propto r^{-\gamma}$ \cite{Rus++03}. The dark matter and baryonic distributions in massive galaxies are individually not power-laws. Yet, a simple power-law radial mass profile has been sufficient to describe both lensing \cite{Gavazzi07, Auger10b} and non-lensing observables such as stellar dynamics \cite{Cappellari16, Derkenne21} and X-ray intensities \cite{Humphrey10}. The total density profile is well approximated by a power-law close to the isothermal $r^-2$, a phenomenon known as the ``bulge--halo conspiracy'' \cite{T+K04, Dutton14}. \cite{Suyu10} allowed departure from the power-law model using pixelated perturbations in the potential and found that large deviations ($>2$\%) from the power-law form were not required.

\begin{figure*}[!t]
\centering\includegraphics[width=0.96\textwidth]{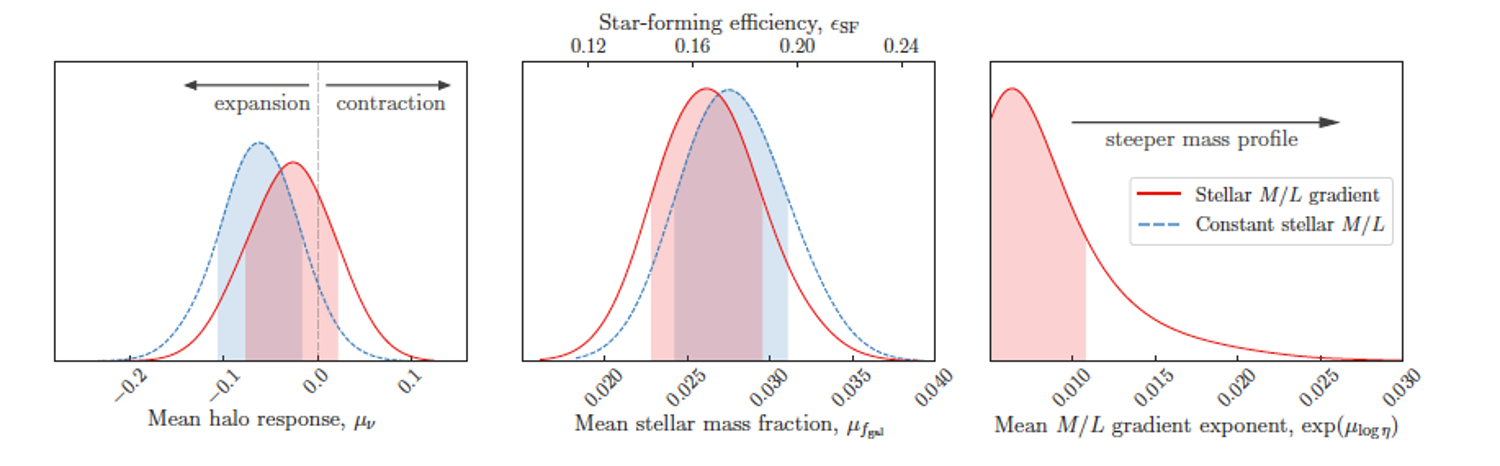}
\centering\includegraphics[width=0.96\textwidth]{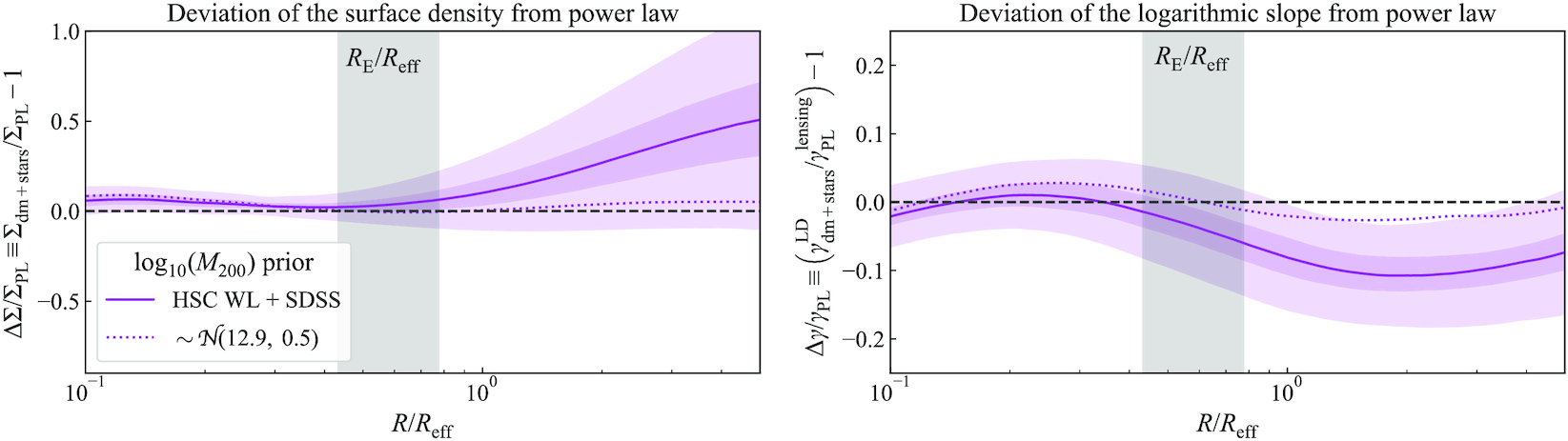}
\caption{Physical interpretation of residual uncertainty allowed by mass-sheet degeneracy on mass density profile.
\cite{Shajib20} modeled a set of non-time-delay lens galaxies with exquisite \textit{HST} images and unresolved stellar velocity dispersion of the deflector fully accounting for the mass-sheet degeneracy and expressed the results as deviations from standard ``composite'' (top row) and power-law mass profiles (bottom row). The standard composite model comprising an NFW \cite{NFW97} dark matter halo and a stellar component with constant mass-to-light ratio is consistent with the data, although a small amount of contraction/expansion of the halo (top left panel) or a small gradient in the mass-to-light ratio (top right panel) cannot be ruled out. Similarly, a power-law mass density profile is consistent with the data, although the data cannot rule out small deviations (purple bands in the bottom panels). See \cite{Shajib20} for more description. When available, additional information -- such as spatially resolved stellar kinematics -- reduces the residual freedom and thus tightens the bounds on \Ho\ when applied to time-delay lenses.}
\label{fig:msd-H0-prof}
\end{figure*}

An alternative choice of simply parametrized mass profile is a composite of the baryonic mass distribution that follows the observed light distribution with a spatially uniform mass-to-light ratio and the dark matter distribution described with a Navarro--Frenk--White profile \cite{Navarro96, Navarro97}. The \ho\ inferred using this composite mass model is reassuringly consistent with that using the simpler power-law model \cite{Millon20}. The composite model with the dark matter distribution described by an NFW profile has also been consistent for a sample of non-time-delay galaxy--galaxy lenses \cite{Dutton14, Shajib21}. These simply parametrized mass models are sufficient to fit the lens imaging data to the noise level and are internally consistent within a few percent for each individual system and within 1\% for the combined \ho\ values from 7 systems \cite{Wong20, Millon20}. However, the true mass distribution can potentially differ from that inferred from lensing data only, owing to the mass-sheet degeneracy \cite{Shajib21}. A physical interpretation for such possible deviation -- small and not required by the data -- is shown in Fig.~\ref{fig:msd-H0-prof}.

To model the deflector light profile, one, two, or three S\'ersic profiles \cite{Sersic68} are used \cite{Suyu10, Wong17, Shajib22}. The quasar host galaxy's light profile can be described on a pixellated grid with a regularization condition \cite{Suyu10b, Vegetti09}. For example, such a pixellated scheme is adopted to reconstruct the source of \examplelens\, illustrated in Fig.~\ref{fig:rxj1131_lens_model}. An alternative parametric approach is to adopt a basis set of a S\'ersic profile and shapelets (i.e., 2D Gauss--Hermite polynomials \cite{Refregier03}), whose amplitudes are determined through linear inversion of the observed image \cite{Birrer15}.

\begin{figure*}[!t]
	\centering
	\includegraphics[scale=.8]{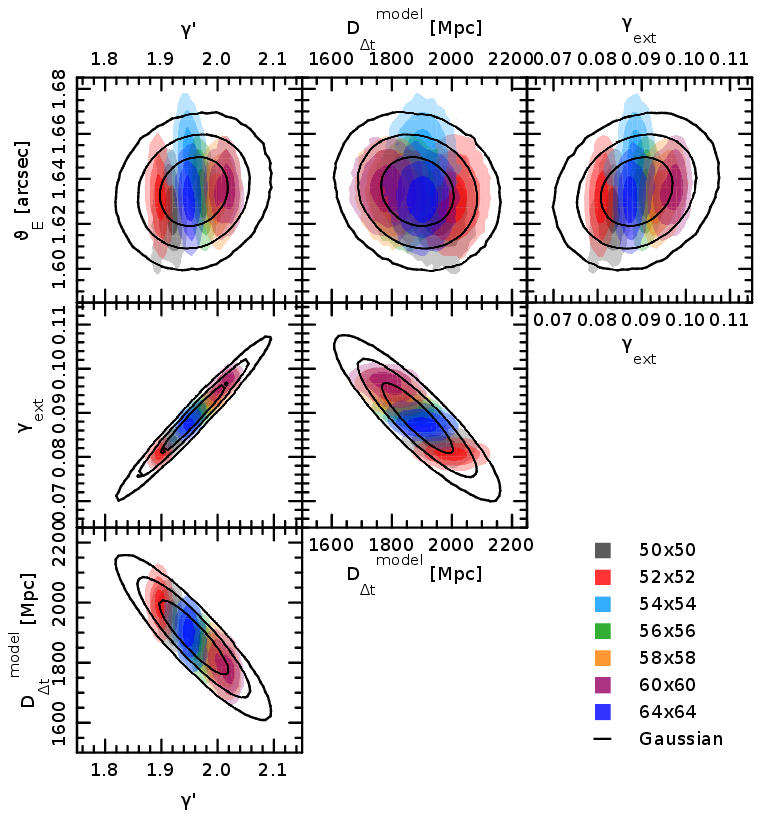}
	\caption{\label{fig:rxj1131_model_corner}
	Combined posterior of the lens model parameters power-law exponent $\gamma$, Einstein radius $\theta_{\rm E}$, and external shear magnitude $\gamma_{\rm ext}$. The illustrated model-predicted time-delay distance $D_{\Delta t}^{\rm model}$ is obtained using a fiducial cosmology from the predicted Fermat potential difference $\Delta \tau_{\rm model}$ from the lens model parameters. Therefore, the illustrated $D_{\Delta t}^{\rm model}$ is simply a reformulation of $\Delta \tau_{\rm model}$, and the measured time delays are still necessary to measure \ho\ using the lens model posteriors. The combined posterior marginalizes over the choice of pixel resolution in the host galaxy reconstruction, where the individual posterior for each choice is illustrated with colored distributions. Figure from \cite{Suyu13}.
	}
\end{figure*}

It is often necessary to make choices in the model settings, for example, the pixel resolution of the reconstructed host galaxy. To avoid any systematic bias and account for modeling uncertainty, different models are constructed by taking a combination of the plausible choices, and then this source of error is marginalized over by combining the posteriors from all these models, {generally weighting by goodness of fit to avoid overcounting relatively poor models}. Fig.~\ref{fig:rxj1131_model_corner} illustrates this marginalization over multiple lens models with differing resolutions in the host galaxy reconstruction. 

\subsection{External convergence from the LOS} \label{sec:los}

The mass structures between the background source and the observer (i.e., galaxies, groups, and clusters) contribute additional lensing effect that can modify the estimated Fermat potential difference and thus the inferred \Ho\ by a few percent. This shift is estimated with the external convergence term $\kappa_{\rm ext}$ in Eq.~\ref{eq:h0_master_equation}. The contributing LOS structures can be grouped into two categories: (i) for lensing effects falling outside the tidal regime\footnote{Tidal regime is when the perturber's gravitational field is smaller than the gradient in the deflection angle field created by the main deflector(s).}, the non-linear lensing contribution needs to be taken into account by directly including their mass distributions in the lens model (described in Section \ref{sec:lens_modeling}), and (ii) for lensing effect falling within the tidal regime, the combined lensing effect can be accounted for with the external convergence term $\kappa_{\rm ext}$ and an external shear term that is already included in the lens modeling. The LOS structures of the first category are typically within $\sim$10$^{\prime\prime}$ from the central deflector. The commonly used criterion to select the perturbers in this category is a ``flexion shift'' threshold \cite{McCully17}. Estimating flexion shift requires photometric redshifts and mass measurements of the LOS galaxies. Spectroscopic redshifts are used over photometric redshifts whenever available. If the LOS galaxies form a group (or cluster), then the group-scale (or cluster-scale) halo also needs to be explicitly modeled if it satisfies the flexion-shift criterion \cite{Keeton04, Momcheva15, Rusu20}. The velocity dispersions of the LOS galaxies are additionally used to infer group or cluster memberships of those galaxies \cite{Sluse19, Buckley-Geer20}. 

The external convergence for the second category of perturbers can be estimated statistically from the number count of galaxies \cite{Fassnacht11} around the lens system (usually within $120^{\prime\prime}$) or using weak lensing effect created by these LOS perturbers on the shapes of background galaxies. The galaxy number counts, either directly or weighted by quantities that correlate with lensing strength, such as projected distance from the central deflector, the perturber's mass, and redshift \cite{Greene13, Rusu17}, are used as summary statistics for the lens environment. LOS cones with matching summary statistics are then selected from cosmological simulations \cite{Hilbert07, Hilbert09}. The external convergence values computed for these simulated cones then provide a probability distribution of the external convergence of the real target (e.g., as obtained by \cite{Suyu10} for \examplelens). The number counts for the observed lens environment are normalized with a large number of control fields for which the photometric data is available at the same quality, and the same is done for the simulated fields. Taking these relative number counts as the summary statistics minimizes the impact of the chosen cosmological parameters in the cosmological simulation. 

From high-quality and wide-field imaging, the distortion in the shapes of background galaxies can be used to constrain the weak lensing shear created by the LOS structures. In the linear regime, this shear uniquely maps to external convergence \cite{Kaiser93}. \cite{Tihhonova18, Tihhonova20} applied this technique to provide alternative estimates of the external convergence $\kappa_{\rm ext}$ for two systems, finding estimates in agreement with those based on galaxy number counts.

\subsection{Stellar kinematics} \label{sec:stellar_kinematics}

\newcommand{\kmps}{km s$^{-1}$ Mpc$^{-1}$}
Stellar kinematics traces the 3D potential of the deflector, whereas lensing traces the 2D projected potential. Thus, in combination, dynamics and lensing can break the degeneracies inherent to each method: the mass-sheet degeneracy in lensing and the mass-anisotropy degeneracy in dynamics. Therefore, stellar kinematics provide the $\lambda_{\rm int}$ term in Eq.~\ref{eq:h0_master_equation}. 

Owing to the angular size and faintness of the deflectors, the most common type of measurement for stellar kinematics is an unresolved (i.e., aperture-integrated) LOS velocity dispersion from long-slit spectra. These spectra are usually taken using large ground-based telescopes, for example, the Keck Observatory and the Very Large Telescope (VLT), in seeing-limited cases in the optical.


The stellar velocity dispersion is usually modeled by solving the Jeans equation \cite{Jeans22, Cappellari20}, which is derived from the collisionless Boltzmann equation \cite{Binney87}. The line of sight velocity dispersion can then be expressed, in the spherical case, as
\newcommand{\dd}{\mathrm{d}}
\begin{equation} \label{eq:los_dispersion}
    \sigma_{\rm los}^2(R) = \frac{2 G}{I(R)} \int_R^\infty \mathcal{K}_\beta \left(\frac{r}{R} \right) \frac{l(r) M(r)} {r}  \dd r,
\end{equation}
where $I(R)$ is the surface brightness distribution of the kinematic tracer in the deflector, $l(r)$ is the 3D luminosity density of the same tracer, $M(r)$ is the 3D enclosed mass, $\mathcal{K}_\beta$ is a function that depends on the anisotropy profile $\beta(r)$ \cite{Mamon05}.

The anisotropy parameter $\beta(r)$ is defined as
\begin{equation}
    \beta(r) \equiv 1 - \frac{\sigma_{\rm t}^2(r)}{\sigma_{\rm r}^2(r)},
\end{equation}
where $\sigma_{\rm t}$ is the tangential component of the velocity dispersion and $\sigma_{\rm r}$ is the radial component. In Eq.~\ref{eq:los_dispersion}, the terms related to the light distribution, that is, $I(R)$ and $l(r)$, are well-constrained from the observed light, albeit with assumptions to deproject from 2D $I(R)$ to 3D $l(r)$. However, there is a degeneracy between the mass distribution giving $M(r)$ and the anisotropy profile giving $\mathcal{K}_{\beta}$, namely the mass--anisotropy degeneracy \cite{Courteau14}. Although combining constraints from lensing and dynamics breaks the mass-sheet and mass--anisotropy degeneracies, the breaking power from unresolved velocity dispersion is limited. In the past, the mass-sheet degeneracy was broken chiefly through the mass profile assumption (i.e., the power-law or composite form), and then the stellar kinematics provided tighter constraint on the mass profile distribution with the mass-anisotropy degeneracy marginalized over \cite{Suyu13}. In such cases, the internal mass-sheet transformation parameter $\lambda_{\rm int}$ (in Eq.~\ref{eq:h0_master_equation}) was effectively set to $\lambda_{\rm int} = 1$. With these assumptions, for \examplelens\ \Ho=$78.3_{-3.3}^{+3.4}$ \hunit\ basend on joint lens modeling constraints from \textit{HST}$+$AO imaging \cite{Suyu14, Chen19}.

\begin{figure*}
	\centering
	\includegraphics[width=\textwidth]{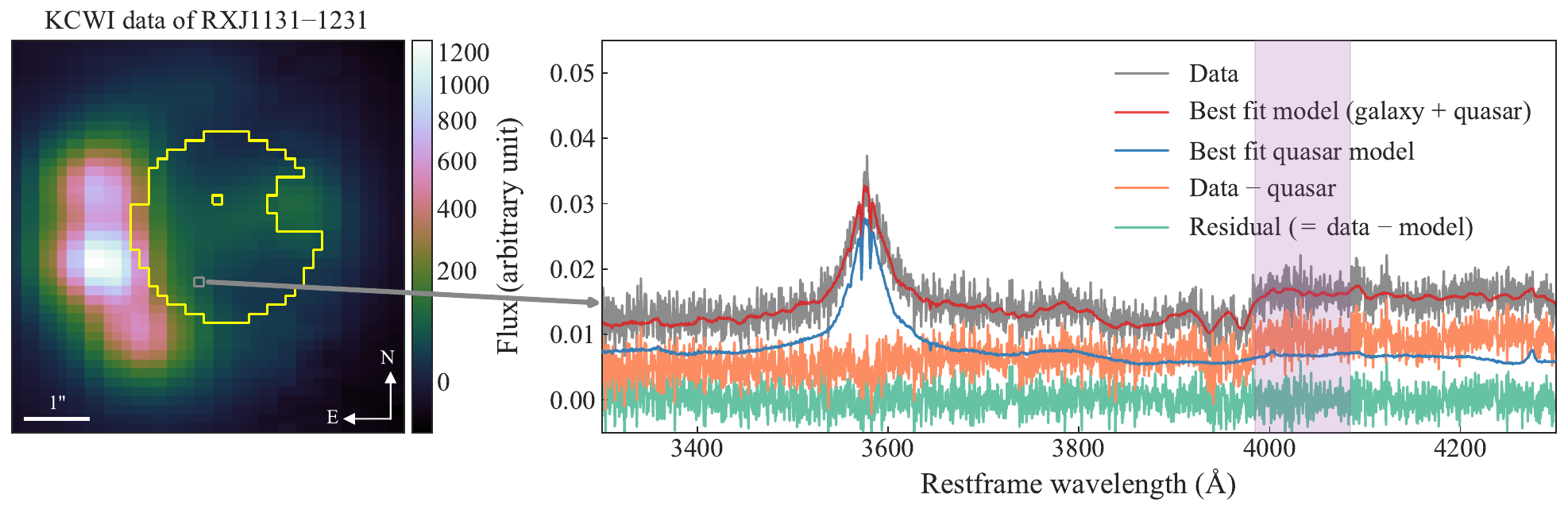}
	\caption{\label{fig:kcwi_spectra_rxj1131}
	Keck/KCWI integral field spectra of \examplelens. \textbf{Left:} 2D representation of the datacube obtained by summing across the wavelength dimension. The yellow contour shows the region within which spectra were extracted to measure the resolved velocity dispersion of the deflector. The grey box marks the pixel for which the observed spectra and model fitting are shown in the right-hand panel. \textbf{Right:} The observed spectra (grey line) and the estimated deflector spectra (orange line) after removing the modeled quasar contribution (blue line). The model for the observed spectra using X-shooter Spectral Library (XSL)  and having fitted the velocity dispersion is shown with the red line. The spatially resolved velocity dispersion map is thus extracted in 41 bins within the yellow contour (Fig.~\ref{fig:resolved_kinematics_rxj1131}). Figure from \cite{Shajib23}.
	}
\end{figure*}

\begin{figure*}
	\centering
	\includegraphics[width=0.8\textwidth]{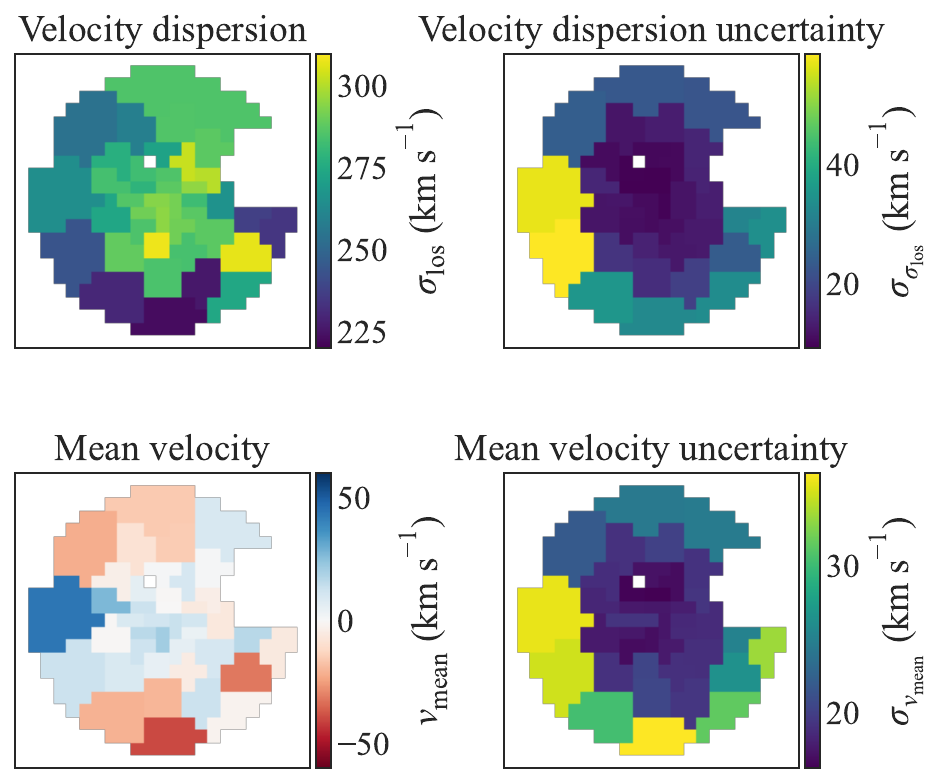}
	\caption{\label{fig:resolved_kinematics_rxj1131} 
	Spatially resolved kinematic maps in 41 bins for \examplelens\ from Keck/KCWI IFU spectra. The top row corresponds to the LOS velocity dispersion, and the bottom row corresponds to the LOS mean velocity. The left column shows the mean values, and the right column shows the uncertainties. Figure from \cite{Shajib23}.
	}
\end{figure*}

Spatially resolved velocity dispersion constrains anisotropy more tightly than the unresolved case. Thus, spatially resolved velocity dispersion is much more effective in simultaneously breaking the above-mentioned degeneracies \cite{Shajib18}. Furthermore, an axisymmetric mass model can be constrained based on the spatial information in the resolved kinematics, allowing one to go beyond simple spherical symmetry in the mass models\cite{Cappellari08}. The oblateness or prolateness of the 3D mass shape can potentially be constrained by the misalignment between the kinematic and photometric major axes \cite{Li18}. The only time-delay lens system with spatially resolved kinematics published so far is \examplelens, based on data from the Keck Cosmic Web Imager (KCWI) on the Keck Telescope (Fig.~ \ref{fig:kcwi_spectra_rxj1131}, \ref{fig:resolved_kinematics_rxj1131}). Obtaining integral field spectra for kinematic measurement has been challenging from the ground since the quasar contribution will contaminate the central deflector's spectra without AO in the seeing limited case as the typical separation between the two is $\sim1^{\prime\prime}$. The kinematic measurement made by \cite{Shajib23} accounted for quasar light contamination in extracting the velocity dispersion of the deflector by forward modeling it. Combining this resolved kinematics with the \textit{HST} imaging, measured time delays, and estimated external convergence yields $H_0 = 77.3_{-7.3}^{+7.1}$ \hunit\ \cite{Shajib23}. This value, obtained by combining resolving kinematics and mass modeling with relaxed assumptions, agrees very well with that obtained from simple parametric mass models, thus corroborating the standard assumptions made in time-delay cosmography \cite{Wong20, Shajib23}.


\begin{figure*}
	\includegraphics[width=1\textwidth]{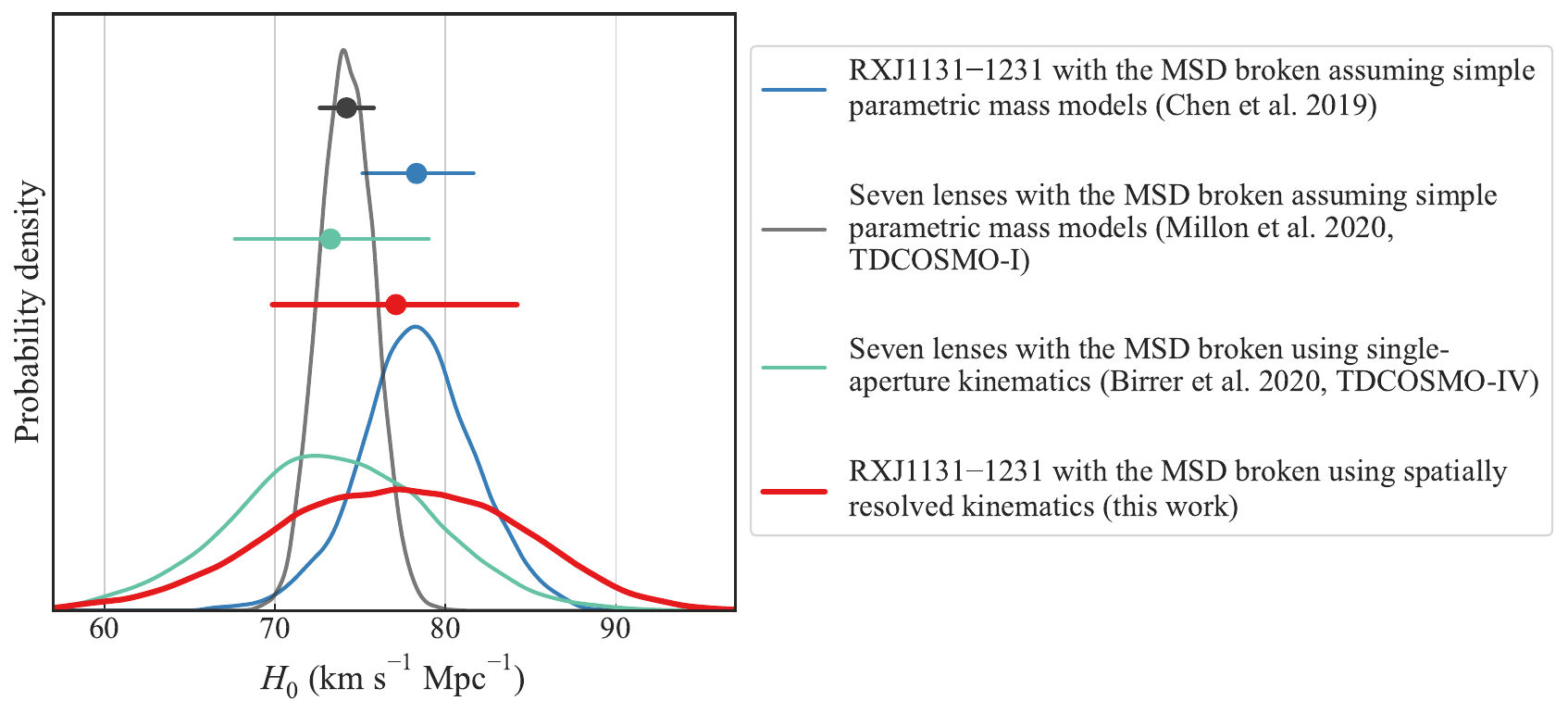}
	\caption{\label{fig:h0_comparison_shajib}
    Comparison of \Ho\ values with various ways to break the mass-sheet degeneracy. For the seven systems analyzed by the TDCOSMO collaboration, the MSD broken by simple parametric assumption on the mass profile (with $\lambda_{\rm int} = 1$ fixed) gives $74.2_{-1.6}^{+1.6}$ \hunit\ (black) \cite{Millon20}, and the MSD broken by unresolved kinematics gives $73.3_{-5.8}^{+5.8}$ \hunit\ (emerald) \cite{Birrer20}. For \examplelens, the MSD broken by simple parametric mass profile assumption gives $78.3_{-3.3}^{+3.4}$ \hunit\ (blue) \cite{Chen19}, and the MSD broken by spatially resolved kinematics gives $77.1_{-7.1}^{+7.3}$ \hunit\ (red) \cite{Shajib23}. Using spatially resolved kinematics for one system gives similar uncertainty on \Ho\ from seven systems with unresolved kinematics, illustrating the superior power of resolved kinematics in breaking the MSD. Fig.~from \cite{Shajib23}.
    }
\end{figure*}


\subsection{Supernova time delay cosmography: the ``Refsdal'' case study}
\label{sec:sn}

\begin{figure*}
	\includegraphics[height=5cm]{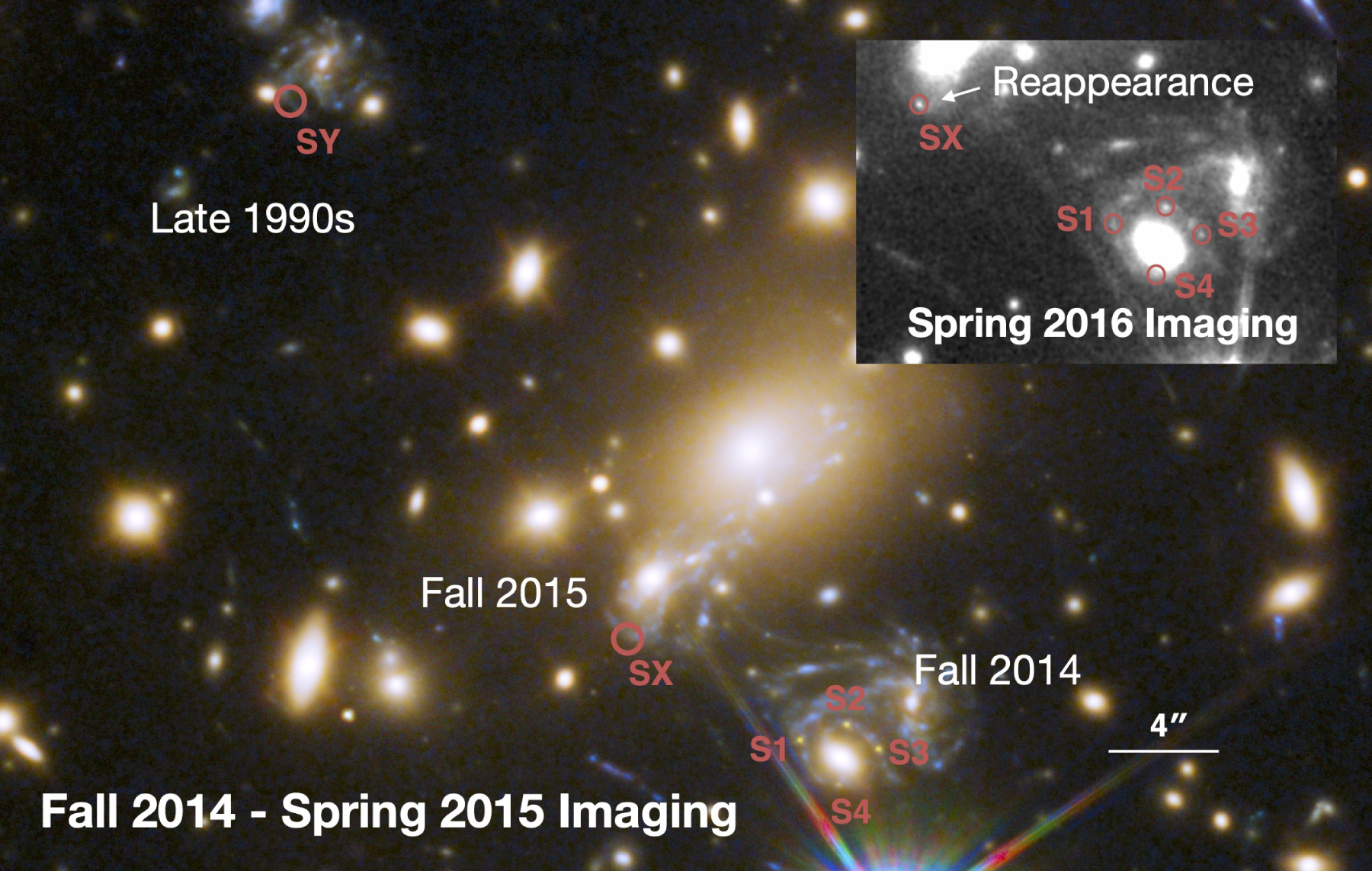}
 \includegraphics[height=5cm]{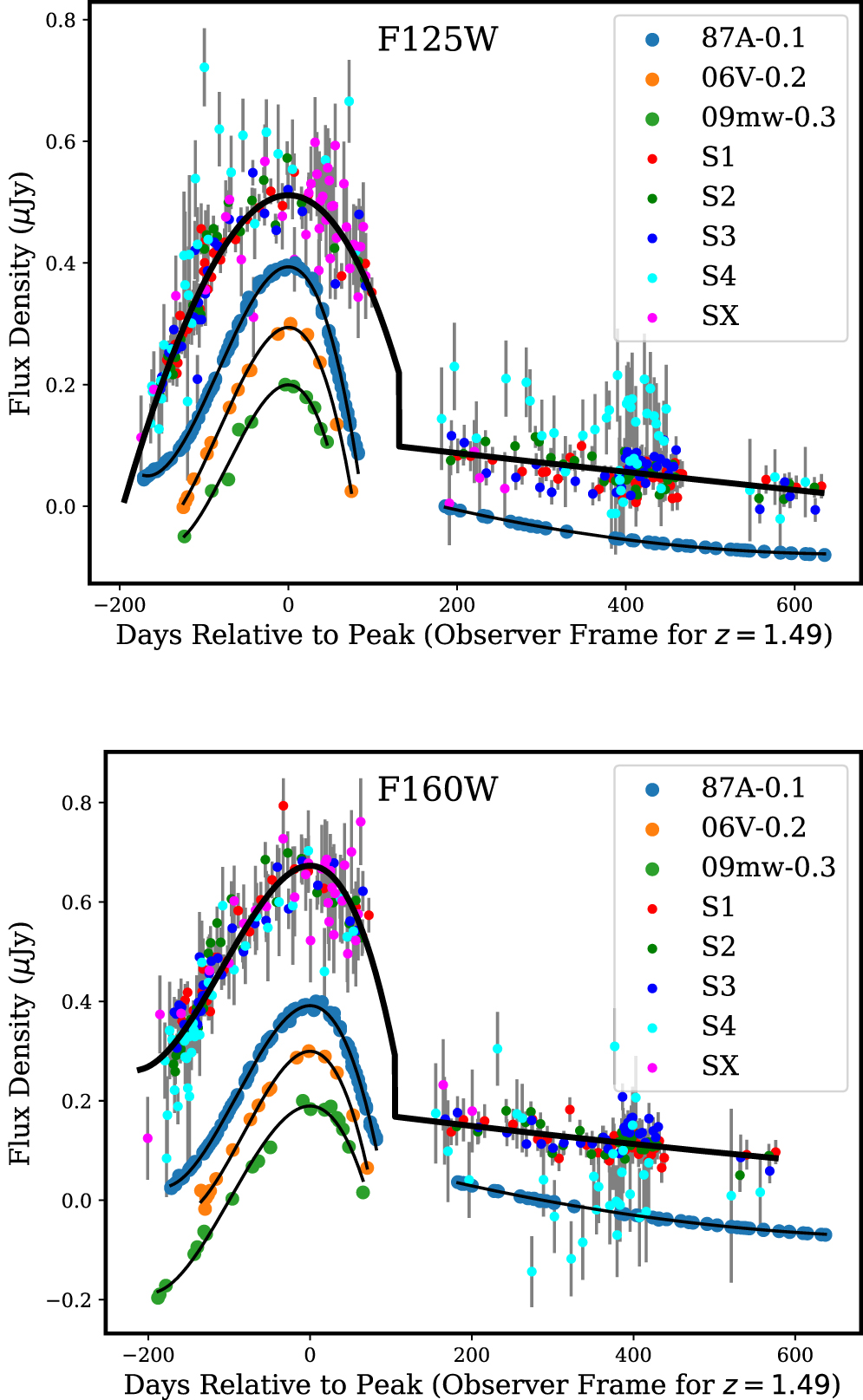}
	\caption{\label{fig:refsdal-td}
    Figure from \cite{Kelly2023a}.
	\textbf{Left:} \textit{HST} images of SN Refsdal, summarizing the time evolution of the phenomenon, including the original discovery of the Einstein Cross in 2014 and the appearance of SX in 2015/2016. \textbf{Right:} light curve of SN Refsdal compared to SN1987A-like light curves and polynomial fits. Figures from \cite{Kelly2023a}.}
\end{figure*}

SN Refsdal (Fig.~\ref{fig:refsdal-td}; \cite{Kelly15}) is the first multiply imaged SN with measured time delays \cite{Rodney2016, Kelly2023a}. It is worth studying this case in some detail as it provides some important lessons for SN time delays, which we expect to be a major contributor to cosmography in the next decade. 

Before going through all the steps leading to the recent measurements of \Ho\ \cite{Kelly2023b}, it is useful to summarize the main differences between this case and the one examined in the previous sections, representing quasars lensed by galaxy-scale deflectors:

\begin{enumerate}
\item The intrinsic light curve of an SN (Fig.~\ref{fig:refsdal-td}) is usually well described by a template or a low-order polynomial. This regular nature simplifies the task of obtaining a high-precision time delay with respect to the stochastic light curves of lensed quasars (Fig.~\ref{fig:rxj1131_light_curve}). Extrinsic effects from the foreground, e.g., microlensing, can affect both cases.
\item The main deflector of SN Refsdal is a cluster of galaxies. Clusters of galaxies are not dynamically relaxed, in contrast to the inner regions of massive elliptical galaxies -- the typical deflectors of galaxy-scale lenses. Thus, cluster lenses are generally significantly more complex to model, both from a lensing and dynamical point of view, with respect to galaxy-scale ones.
\item The caustics of clusters of galaxies cover a much larger solid angle on the sky than those of galaxies. Therefore, tens or even hundreds \cite{Bergamini2023} of multiply imaged sources can be used to constrain the lens model in a cluster. There is generally one multiply imaged source for galaxy-scale lenses, although systems with up to a handful of them have been found \cite{Gavazzi08, Shajib19}. Note that having several families of multiply imaged sources at different redshifts helps mitigate the mass-sheet degeneracy \cite{BLS04, Grillo2018, Grillo2020}.
\end{enumerate}

\begin{figure*}
	\includegraphics[height=4cm]{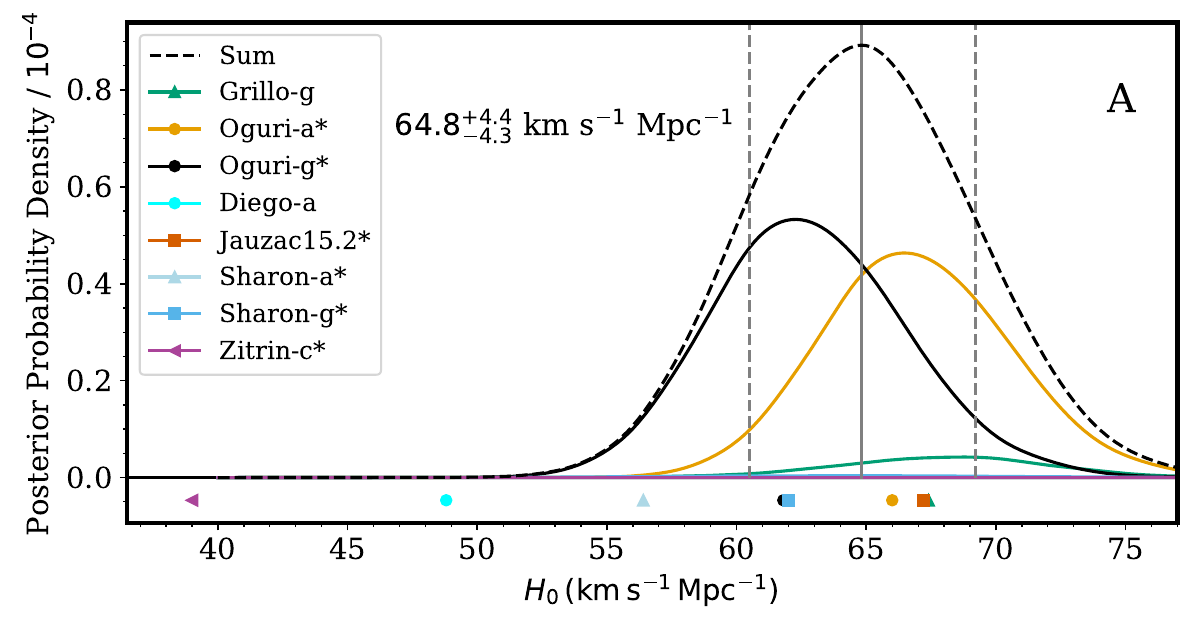}
 \includegraphics[height=4cm]{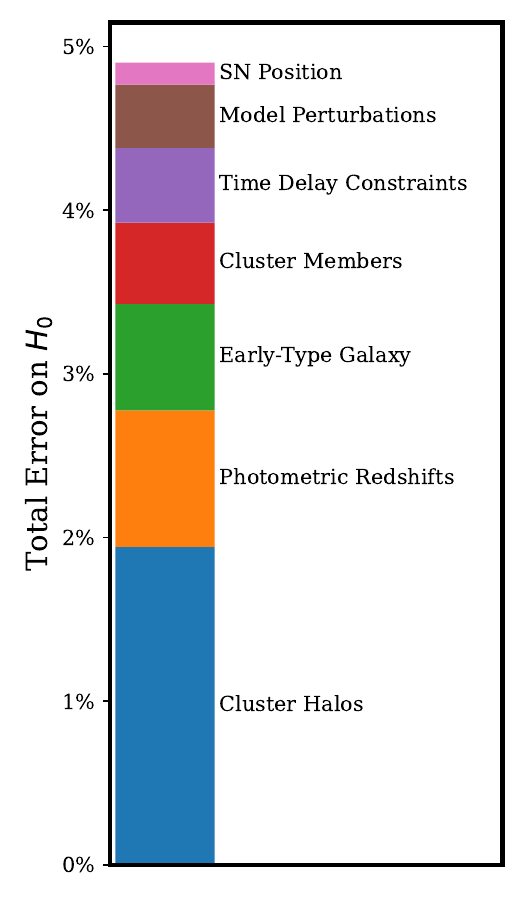}
	\caption{\label{fig:refsdal-H0} Left: unblinded posterior distribution function of H$_0$ based on supernova Refsdal. The lens models have been weighted according to their agreement with the H$_0$-independent observables, such as image positions and magnification ratios. Right: error budget on H$_0$ measurement.
    Figures from \cite{Kelly2023b}.
	}
\end{figure*}

SN Refsdal was first discovered \cite{Kelly15} as an Einstein cross in images taken as part of the Grism Lens-Amplified Survey from Space (GLASS) program \cite{Treu15}. Although the time delays between the cross images were expected to be of order days/week (and thus not very useful for cosmography) because of the symmetry and separation of the configuration, it was immediately recognized \cite{Kelly15, Oguri15, Sharon15} that one additional and more distant image (SX) would re-appear some time in the near future, with great potential for cosmography. 

Considerable effort went into predicting SX's timing, position, and brightness using updated lens models based on extensive spectroscopy \cite{Treu16b, Grillo16} before it appeared in the sky. Indeed, SX appeared as predicted \cite{Kelly15}, helping to build confidence in the models. The good agreement was not a foregone conclusion, considering the complexity of the cluster lens models.

The first estimate of the magnification ratio and of the long time delay (SX to the cross; the one with the most cosmographic constraining power) was used by one team to produce an estimate of \Ho\ \cite{Vega-Ferrero2018}. That study highlighted the spread between the different cluster lens models. The spread between all the possible models is not surprising, considering that many of the lensing codes employed were not designed for precision cosmology and, therefore, did not have the necessary numerical precision and resolution. Pixellated models and models based on heavily regularized basis sets do not have by design the angular resolution needed for time-delay cosmography. The resolution requirement can be understood in terms of astrometric precision. Measuring the Hubble constant to a few percent precision usually requires knowing how the Fermat potential (and image positions) vary over tens of milliarcseconds \cite{Birrer19}. Furthermore, many of the early models did not make full use of all the available information, e.g., cluster membership, stellar velocity dispersions of cluster members, and spectroscopic redshift of multiple images. A thorough discussion of the sources of uncertainty is given by \cite{Grillo2020}.

After the re-appearance, a major effort was undertaken to obtain a blind, high-precision (1.5\%) measurement of the time delay \cite{Kelly2023a} and fold it in with existing lens models to obtain a blind measurement of \Ho\ \cite{Kelly2023b}.  It is crucial to stress that all the analysis choices were made blindly with respect to \Ho\ (the time delay was kept blind throughout the analysis, for example) and that the analysis was not modified after unblinding (with the exception of the correction of a mistake in sign convention on magnifications).  This is a crucial step to prevent ``experimenter bias,'' and we advocate that every cosmological measurement should be carried out blindly. 

As highlighted by \cite{Vega-Ferrero2018}, the spread between lens models is clearly significant. It should be noted, however, that they include in their analysis models that were later discovered to be affected by substantial numerical errors \cite{Kelly2023b}, and therefore their spread was overestimated.  Two options were considered by \cite{Kelly2023b} to account for the spread between models. First, it was decided to consider only the two models that are based on codes designed to do high-precision cosmography, \textsc{glee} \cite{S+H10} and \textsc{glafic} \cite{Oguri10}, weighted by their agreement with observables independent of \Ho, yielding $H_0=66.6^{+4.1}_{-3.3}$ \hunit. {If one had preferred to do the straight average of the two models selected prior to unblinding to be the most accurate, the result would have changed negligibly:} combining the \textsc{glee} and \textsc{glafic} models with equal weight yields $H_0=67.2^{+4.1}_{-3.7}$ \hunit, consistent within the errors. Alternatively, an analysis considering all models was run, again weighted by the same scheme, finding $H_0=64.8^{+4.3}_{-4.3}$ \hunit. The results are very similar between the two schemes because the \textsc{glee} and \textsc{glafic} models provide, by far, the best match to the observables. {The good agreement between the two options is partly due to the fact that the \textsc{glee} and \textsc{glafic} models are by far the ones that best match the observables. This is not surprising, considering that some of the other models were not designed for precision cosmography, as discussed by \cite{Treu2016} and \cite{Kelly2023a}.   Figure~2 in \cite{Kelly2023a} shows that the models with the least weights (Zitrin and Diego) have very broad tails to significantly lower \Ho, while Jauzac and Sharon yield results comparable to Grillo and Oguri. }

The measurement from SN Refsdal \cite{Kelly2023a} is not precise enough to help solve the ``Hubble tension.'' Given the uncertainties \cite{Grillo2020, Kelly2023a}, it is consistent with both \textit{Planck} and the local distance ladder method \cite{Riess2022}. The latter is only 1.5$\sigma$ away from the best fit, thus consistent with the SN Refsdal analysis.

The important lessons from SN Refsdal are two. First, the blind efforts succeeded to the level that only the most optimistic practitioners of cluster lens modeling would have expected. SX appeared exactly where and when it was predicted to be, and the inferred value of \Ho\ is perfectly consistent with those of other methods. It did not have to be this way, considering the complexity of the mass distribution in \macs. SX could have appeared elsewhere or at a different time, and \Ho\ could have turned out to be 30 or 100 \hunit. {This blind prediction helps build confidence in the models. Although it is clearly impossible to prove that a better model does not exist, the best models appear to be sufficiently precise and accurate, to the extent that we can probe them empirically.} 
Second, 6\% is a very small uncertainty for an absolute distance measurement from a single system, and compelling arguments show that it is not significantly underestimated \cite{Grillo2020, Kelly2023b} when the data quality is as high as in this case. {We stress that high-quality data is vital to obtain this kind of precision. The model from \cite{Grillo2020} is constrained by 89 spectroscopically confirmed multiple images arising from 28 distinct sources. Most clusters to date do not have the same level of empirical constraints \cite{Oguri2023}, although the situation is rapidly improving \cite{Bergamini2023}.}  A simple $\sqrt{N}$ scaling suggests that ten systems similar to \macs\ will suffice to reach $\sim 2$\% precision and thus contribute to solving the Hubble tension if no yet-to-be-discovered systematic floor arises.


\section{Conclusions and future outlook}
\label{sec:outlook}

\begin{figure}[t!]
\centering
\includegraphics[width=0.9\textwidth]{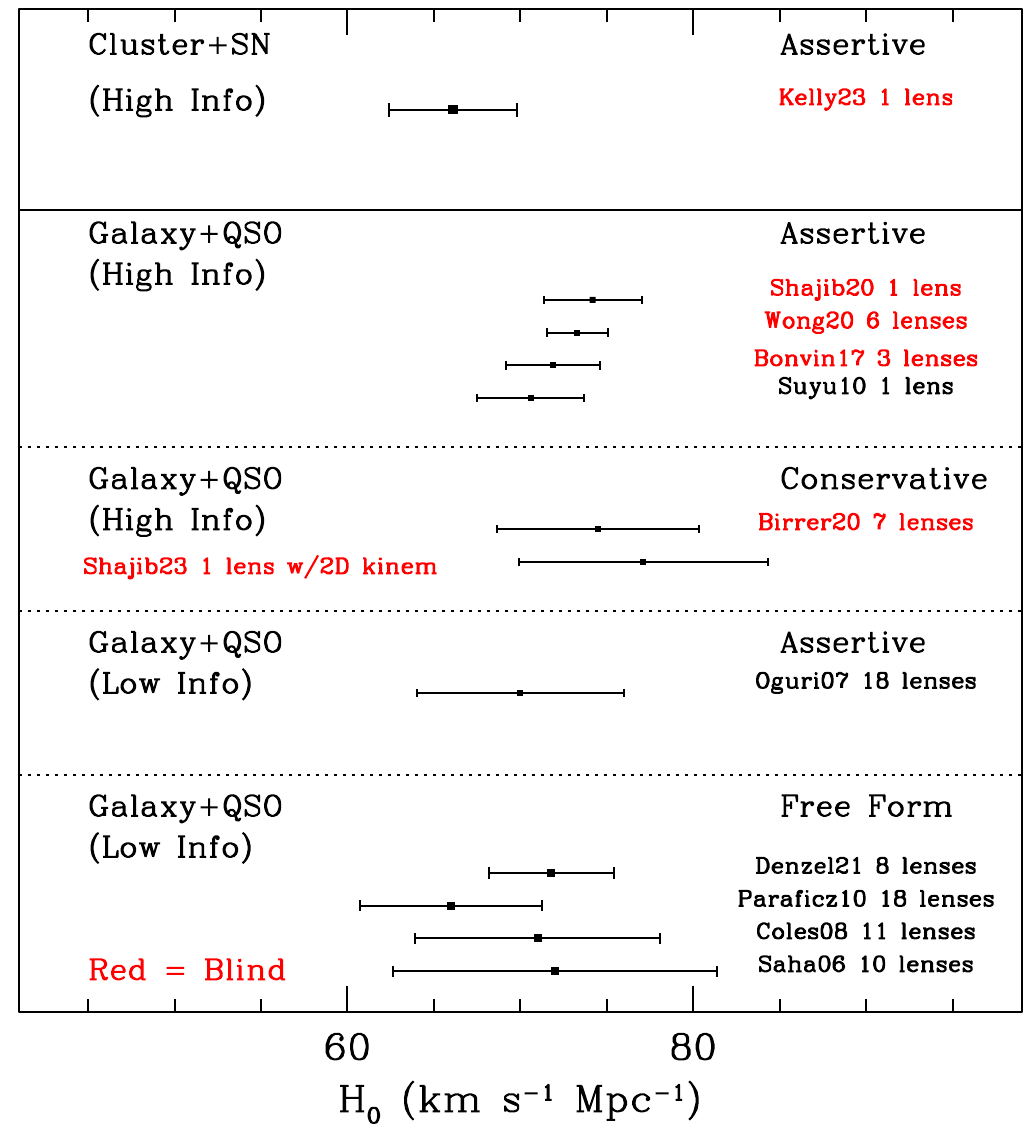}
\caption{Comparison of \Ho\ measurements based on time-delay cosmography, in $\Lambda$CDM cosmology. The measurements are grouped by: i) the lensing configuration (galaxy+QSO vs. cluster+SN). ii) Assumptions on the mass distribution of the main deflector, "assertive" and "conservative" for simply parametrized models or "free form" for pixellated models. iii) Amount of information used per lens; in the case of a galaxy-scale lens, "low info" utilizes quasar positions and time delays, "high info" adds the extended surface brightness distribution of the multiple images of the quasar host galaxy, stellar kinematics of the main deflector, and number counts or weak lensing to estimate the line of sight convergence. For the cluster+SN case, we define it as high info due to the large number of spectroscopically confirmed multiply-images and cluster members, and we define it as assertive because the \textsc{glee} and \textsc{glafic} models used for this measurement are based on simply parametrized forms. We give the reference and the number of time delay lenses for each measurement. The measurements shown in red have been blinded to prevent experimenter bias. The figure is updated from \cite{Treu22}.}
\label{fig:plotH0}
\end{figure}

The two recent case studies we chose to highlight represent genuine breakthroughs. 
We will now discuss them in the context of other measurements to give a sense of the landscape (a selection of measurements is summarized in Figure~\ref{fig:plotH0}).

\examplelens\ is the first galaxy-scale lens with time delay and spatially resolved kinematics. The exquisite data for this system enabled \cite{Shajib23} to reach 9\% precision from a single lens, using what we call ``conservative'' assumptions about the mass distribution, i.e., allowing the mass-sheet degeneracy to be only constrained by kinematic data, and thus allowing maximum uncertainty on \Ho. This is remarkable, as the precision is comparable to what was obtained by \cite{Birrer20} with seven lenses for which only unresolved kinematics were available. With the higher angular resolution attainable with JWST or with future instruments with adaptive optics, we can expect the precision per system to reach $\sim 4$\% \cite{Yildirim21}, better than that was achieved in the past for individual systems by using the more ``assertive'' approach of breaking the mass-sheet degeneracy by imposing a functional form for the mass density profile \cite{Shajib20}. With this kind of precision, reaching the 2\% precision, which was previously achieved under ``assertive'' assumptions by \cite{Wong20, Millon20}, seems within reach under ``conservative'' assumptions with the systems known today \cite{Birrer21}.``Free form'' models, that is, the ones in which the surface mass density or potential of the main deflector is rendered as pixels (see \cite{Denzel21, Paraficz2010, Coles2008, Saha2006}, should be able to obtain similar precision if they can be constrained by the full, high-information content from data with state-of-the-art quality. This has not been achieved yet but should be attainable. Along the way, work will need to continue in order to uncover and mitigate new sources of systematic errors, including those arising from selection effects \cite{Collett16}.

The success of \macs\ paves the way for lensed SNe to become a major contributor to time-delay cosmography. A single system with excellent data quality, modeled in an ``assertive'' way, obtains a 6\% precision on \Ho, comparable to the average time-delay quasars of those analyzed by \cite{Wong20}.  With major synoptic surveys such as \textit{Euclid} and the Vera C.~Rubin Observatory's Legacy Survey of Space and Time (LSST) about to begin, we are confident that many such systems will be discovered in the coming decade \cite{Oguri10}. Hopefully, samples of lensed SN will soon reach comparable precision (or better!) to lensed quasars, thus providing a vital sanity check and increased overall precision with respect to quasar-only forecasts such as those presented by \cite{Treu22}, thus getting us closer to solving the ``Hubble Tension.'' 

\begin{acknowledgement}
TT and AJS thank their many colleagues and friends working in the field of time delay cosmography, without whom the progress described here would have been impossible. We thank Fred Courbin, Pat Kelly, Veronica Motta, and Paul Schechter for their constructive comments on an early draft of this manuscript.
TT gratefully acknowledges support by the National Science Foundation, the National Aeronautics and Space Administration, the Packard Foundation, and the Moore Foundation. 
Support for this work was provided by NASA through the NASA Hubble Fellowship grant HST-HF2-51492 awarded to AJS by the Space Telescope Science Institute, which is operated by the Association of Universities for Research in Astronomy, Inc., for NASA, under contract NAS5-26555.
\end{acknowledgement}

\clearpage




\printbibliography


\end{document}